\documentclass[
 reprint,
 amsmath,amssymb,
 aps, prb,
 showpacs
]{revtex4}

\usepackage{amsmath,amssymb}
\usepackage{textcomp}
\usepackage{float}
\usepackage{latexsym}
\usepackage{graphicx}
\usepackage[usenames]{color}
\usepackage{hyperref}
\usepackage{fmtcount}
\usepackage{cleveref}
\Crefname{figure}{FIG.}{FIGs.}
\usepackage{dcolumn}% Align table columns on decimal point
\usepackage{bm}% bold math
\usepackage{epstopdf}

\newcommand{\bear}{\begin{eqnarray}}
\newcommand{\eear}{\end{eqnarray}}

\expandafter\def\expandafter\UrlBreaks\expandafter{\UrlBreaks
  \do\a\do\b\do\c\do\d\do\e\do\f\do\g\do\h\do\i\do\j%
  \do\k\do\l\do\m\do\n\do\o\do\p\do\q\do\r\do\s\do\t%
  \do\u\do\v\do\w\do\x\do\y\do\z\do\A\do\B\do\C\do\D%
  \do\E\do\F\do\G\do\H\do\I\do\J\do\K\do\L\do\M\do\N%
  \do\O\do\P\do\Q\do\R\do\S\do\T\do\U\do\V\do\W\do\X%
  \do\Y\do\Z}

\begin{document}

%%%%%%%%%%%%%%%%%%%%%%%%%%%%%%%%%%
\title{An undergraduate laboratory experiment on measuring the velocity of light with a do-it-yourself catastrophic machine}
%%%%%%%%%%%%%%%%%%%%%%%%%%%%%%%%%%

\author{Todor~M.~Mishonov}
\email[E-mail: ]{mishonov@gmail.com}
\affiliation{Department of Theoretical Physics, Faculty of Physics,\\
St.~Clement of Ohrid University at Sofia,\\
5 James Bourchier Blvd., BG-1164 Sofia, Bulgaria}

\author{Albert~M.~Varonov}
\email[E-mail: ]{avaronov@phys.uni-sofia.bg}
\affiliation{Department of Theoretical Physics, Faculty of Physics,\\
St.~Clement of Ohrid University at Sofia,\\
5 James Bourchier Blvd., BG-1164 Sofia, Bulgaria}

\author{Dejan~D.~Maskimovski}
\email[E-mail: ]{dejan_maksimovski@yahoo.com}
\affiliation{Private Yahya Kemal High School Skopje -- Butel,\\
Butelska bb. Blvd., MKD-1000 Skopje, Macedonia}

\author{Stojan~G.~Manolev}
\email[E-mail: ]{manolest@yahoo.com}
\affiliation{Middle School Goce Delchev,\\
Purvomaiska str. 3, MKD-2460 Valandovo, R.~Macedonia}

\author{Vassil~N.~Gourev}
\email[E-mail: ]{gourev@phys.uni-sofia.bg}
\affiliation{Department of Atomic Physics, Faculty of Physics,\\
 St.~Clement of Ohrid University at Sofia,\\
5 James Bourchier Blvd., BG-1164 Sofia, Bulgaria}

\author{Vasil~G.~Yordanov}
\email[E-mail: ]{vasil.yordanov@gmail.com}
\affiliation{Department of Atomic Physics, Faculty of Physics,\\
 St.~Clement of Ohrid University at Sofia,\\
5 James Bourchier Blvd., BG-1164 Sofia, Bulgaria}

%%%%%%%%%%%%%%%%%%%%%%%%%%%%%%%%%%

%%%%%%%%%%%%%%%%%%%%%%%%%%%%%%%%%%

%%%%%%%%%%%%%%%%%%%%%%%%%%%%%%%%%%
\begin{abstract}
An experimental setup for electrostatic measurement of $\varepsilon_0$, separated magneto-static measurement of $\mu_0$ and determination of the velocity of light $c=1/\sqrt{\varepsilon_0 \mu_0}$  according to Maxwell theory with percent accuracy is described.
No forces are measured with the experimental setup therefore there is no need of a scale and the experiment price less than \pounds20 is mainly due to the batteries used.
Multiplied 137~times, this experimental setup was given at the fourth open international Experimental Physics Olympiad (EPO4) and a dozen high school students did very well.
This article, however, focuses on the catastrophe theory, which is the basis of the methodology.
\end{abstract}
%%%%%%%%%%%%%%%%%%%%%%%%%%%%%%%%%%

\maketitle

%%%%%%%%%%%%%%%%%%%%%%%%%%%%%%%%%%
%
%%% Experimental setup
%
%%%%%%%%%%%%%%%%%%%%%%%%%%%%%%%%%%

\section{Experimental setup description}

%%%%%%%%%%%%%%%%%%

\subsection{Electrostatic experiment}

Imagine a parallel plate capacitor.
One of the plates is a suspended pendulum at $x_1=0$.
The other plate is fixed on a movable block at $x$.
When a voltage $\mathcal{E}$ is applied to the capacitor, the pendulum is shifted towards the other plate to a distance $y$.
The distance between the parallel disk-shaped plates becomes
$z=x-y > 0$.
The pendulum length $L_\mathrm{e} \approx 54\,\mathrm{cm}$ is much larger than the shift and for the restoring gravitational force we have approximately Hooke's law
%%%
\begin{equation}
F_\mathrm{g} = -\frac{\mathrm{d}}{\mathrm{d}z} \left (U_g(z) = m_\mathrm{e}g \sqrt{L_\mathrm{e}^2 - y^2} \right) \approx m_\mathrm{e}g \frac{x-z}{L_\mathrm{e}}.
\nonumber
\end{equation}
%%%
For brevity in one and the same expression we introduced 
the potential energy $U_g$, the force $F_\mathrm{g}$ calculated as its derivative and gave the approximate expression used in the present work.
We use $D_\mathrm{e}=2 R_\mathrm{e} = 54\,\mathrm{mm}$ diameter aluminum plates punched according to EC standard jar caps with mass $m_\mathrm{e}=1.14\,\mathrm{g}.$

Here we are not going to rewrite a textbook on electrodynamics,
our purpose is to give a concise reference to many formulae for the force which can be found in the literature.
The electrostatic force $F_\mathrm{e}$ is the gradient of the effective potential energies\cite{LL} defined in the parentheses below
%%%
\begin{equation}
F_\mathrm{e}(z) = -\frac{\mathrm{d}}{\mathrm{d}z} \left. 
\left ( U_\mathcal{_E}(z,\mathcal{E}) 
= -\frac{1}{2} C(z) \mathcal{E}^2 \right )\right|_{\mathcal{E}=\mathrm{const}} =
-\frac{\mathrm{d}}{\mathrm{d}z} \left. \left (U_{_Q}(z,Q) \equiv  \frac{1}{2} \frac{Q^2}{C(z)} \right )\right|_{Q=\mathrm{const}} =
 \frac{1}{2} \left(\mathcal{E}^2 = \frac{Q^2}{C^2} \right)
\frac{\mathrm{d}}{\mathrm{d}z} C(z) ,
\nonumber
\end{equation}
%%%
where $Q=C\mathcal{E}$ is the capacitor charge. 
The concise expression above could be described in six different rows with one page text between them thus losing the transparent physical meaning. 
Nevertheless, some people prefer the horrible pleonasm of the detailed sequential description.
Only referred formulae are numbered because reference to a numbered formula is like GOTO operator in the programming. 

The capacity can be calculated as the energy of the electric field 
$\mathbf{E}(\mathbf{x}) = -\nabla \varphi $
%%%
\begin{equation}
Q=\varepsilon_0 \oint \mathbf{E} \cdot \mathrm{d} \mathbf{S},\qquad
U_{_Q} = \frac{1}{2} \frac{Q^2}C =  \int \mathrm{d}^3 x \frac{1}{2} \varepsilon_0 E^2,\quad U_\mathcal{_E}\equiv U_{_Q}-Q\mathcal{E}=-U_{_Q},\quad \Delta\varphi=0,
\quad \varphi_1=0,\quad \varphi_2=\mathcal{E},
\nonumber
\end{equation}
where 
the first surface integration is around one of the plates of the capacitor
and the second volume integration is over the whole 3-dimensional space.
The electrostatic energy $U_{Q}$ calculated as integral of the energy density on the whole space is a positive variable.
The electric potential on the electrodes of the capacitor is constant while 
in free space it is a harmonic function.
Different expressions for the force 
%%%
\begin{equation}
F_e=-\left(\frac{\partial U_{_Q}}{\partial z}\right)_{\!\! Q}
=-\left(\frac{\partial U_\mathcal{_E}}{\partial z}
\right)_{\!\mathcal{E}}, 
\qquad
\mathcal{E}=+\left(\frac{\partial U_{_Q}}{\partial Q}\right)_{\!z}=\frac{Q}{C},
\qquad
Q=-\left(\frac{\partial U_\mathcal{_E}}{\partial \mathcal{E}}
\right)_{\!z}=C\mathcal{E}
\nonumber
\end{equation}
are convenient for different type experiments at fixed voltage $\mathcal{E}$ or at fixed charge $Q$. 
Both type experiment can be done in the described experimental set-up.

One more intuitive point of view for the effective potential $U_\mathcal{_E}$ is to consider as a \textit{Gedanken Experiment} 
parallel switching of one big capacitor $C_0\gg C(z) $ charged by a voltage $\mathcal{E}$ at $z=\infty$, when $C(z=\infty)=0.$ 
For the charge of the big capacitor we have $Q_0=C_0 \mathcal{E}$ and this charge is conserved.
After opening of the parentheses, the total electrostatic energy
reads
%%%
\begin{equation}
U_\mathrm{tot}(z)=\frac{Q^2}{2C(z)}+\frac{(Q_0-Q)^2}{2C_0}
=U_\mathcal{_E}(z,\mathcal{E})+\frac12 C_0\mathcal{E}^2
+\frac{Q^2}{2C_0}.
\nonumber
\end{equation}
%%%
The last term is negligible $1/C_0\ll 1/C(z)$, the middle is a constant irrelevant with respect to $z$ differentiation, and again we arrive at  
$U_\mathcal{_E}(z,\mathcal{E})=-\frac12C(z)\mathcal{E}^2$ using a charge reservoir as an auxiliary construction. 
The effective potential $U_\mathcal{_E}$ is negative because it describes the energy of an open system including the energy spent by external voltage source to keep voltage constant. The second derivation is more understandable for students not familiar with the thermodynamic style of writing the derivatives.

The position $z$ of the shifted pendulum by the electric field is determined by the minimum of the total energy
%%%
\begin{equation}
U_\mathrm{e}(z)=U_\mathrm{g}(z)+U_\mathcal{_E}(z).
\nonumber
\end{equation}
%%%
The experiment is conducted at DC voltage,
but if AC current is used $\mathcal{E}$ is the RMS value.

The forces are balanced in equilibrium and the total force
%%%
\begin{equation}
F(z_0) = - \frac{\mathrm{d}}{\mathrm{d}z} U_\mathrm{e}(z=z_0)=0.
\nonumber
\end{equation}
%%%
The equilibrium is stable if the potential energy second derivative is positive
%%%
\begin{equation}
k_\mathrm{e}(z) \equiv - \frac{\mathrm{d}^2}{\mathrm{d}z^2}  U_\mathrm{e}(z=z_0) >0.
\nonumber
\end{equation}
%%%
Then, for small deviations from equilibrium, we again have Hooke's law for the force
%%%
\begin{equation}
F(z) \approx -(z-z_0) k_\mathrm{e}(z_0) 
\nonumber
\end{equation}
%%%
and the oscillations frequency of the pendulum
%%%
\begin{equation}
\omega = \sqrt{k_\mathrm{e}(z_0)/m_\mathrm{e}},
\nonumber
\end{equation}
%%%
if the friction force is negligible.
We can describe the experiment now.

At fixed voltage $\mathcal{E}=\mathrm{const},$ after waiting for the oscillations to attenuate, we move the block very slowly towards the pendulum, decreasing the control parameter $x$.
We can note that the oscillations frequency also decreases and their period $T=2\pi/\omega(z_0)$ increases threateningly, and that critical slowing down is the precursor of the stability loss.

The system loses stability $F(z_\mathrm{e}) = 0$ and $k_\mathrm{e}(z_\mathrm{e})=0$ at some critical value $x_\mathrm{e}$ of the control parameter, and evanescent perturbations of the pendulum suddenly swing it towards the block.
Such a leap or a catastrophic change of the state of the systems at a slight variation of the control parameter is systematically described by the catastrophe theory. In our case, the potential energy at $x_\mathrm{e}$ has an inflection point
%%%
\begin{equation}
\left. \mathrm{d}_z U_\mathrm{e}(z,x_\mathrm{e})\right|_{z=z_0}=0 \qquad \& \qquad \mathrm{d}_z^2 \left. U_\mathrm{e}(z,x_\mathrm{e})\right|_{z=z_0}=0, \qquad  \mathrm{d}_z \equiv \frac{ \mathrm{d}}{\mathrm{d} z}.
\label{einfl}
\nonumber
\end{equation}
%%%
If we substitute in this system the Helmholtz formula for a round capacitor,
see for example 8th volume of the Landau-Lifshitz Course of theoretical physics\cite{LL} 
%%%
\begin{equation}
C(z)=4 \pi \varepsilon_0 \left ( \frac{S}{4 \pi z} + \frac{l}{8 \pi^2} \ln\frac{\sqrt{S}}{z} + \overline{C}_\mathrm{H} \right ),
\qquad 
S=\pi R_\mathrm{e}^2, \qquad l=2 \pi R_\mathrm{e},
\qquad \overline{C}_\mathrm{H}=\ln(16 \sqrt{\pi}) - 1= 2.34495,
\nonumber
\end{equation}
%%%
after some algebra we get
%%%
\begin{equation}
\varepsilon_0 \mathcal{E}^2 \approx 
\mathcal{F}_\mathrm{e}\equiv
\frac{32}{27 \pi} \frac{m_\mathrm{e}g}{L_\mathrm{e} D_\mathrm{e}^2} \left [ 1 - f_\mathrm{e}(x_\mathrm{e}/D_\mathrm{e}) \right ] x_\mathrm{e}^3, \qquad f_\mathrm{e}\approx\frac{4}{3 \pi} \frac{x_\mathrm{e}}{D_\mathrm{e}}>0. 
\label{eps0}
\nonumber
\end{equation}
%%%
There are only measurable quantities on the right and electric ones on the left;
the dimensionality of this equation is force.

We repeat the experiment for different voltages, for instance 100, 200, \dots, 800~V provided by 23A batteries placed in plastic tubes 
($16 \times 12\,\mathrm{V}\approx 200\,\mathrm{V}$). 
This is a safety measure for the high school students participating in the Olympiad, while a standard voltage source could be used in a university student laboratory.
After the plates stick to each other, the capacitor is short-circuited and the distance $x_\mathrm{e}$ is carefully measured with a ruler with 0.5~mm accuracy.
The proportionality coefficient  $\varepsilon_0$ at the different voltages $\mathcal{E}$ is determined via the standard method for linear regression. 
And the experimental points are fitted in 
($\mathcal{F}_\mathrm{e}$, $\mathcal{E}^2 $)~plane with a straight line with high correlation coefficient.\cite{EPO4}
%%%%%%%%%%%%%%%%%%

\subsection{Magneto-static experiment}

The magneto-static experiment is practically identical to the electrostatic one.
The attracting metal disks are substituted with attracting coils with diameter $D_\mathrm{m}=2R_\mathrm{m}=65\,\mathrm{mm}$ and N=50~turns of
80~$\mu$m Cu wire
which parallel currents $I$ flow through.
The most sensitive range of the used multimeters is 200~mA therefore the currents for the magneto-static experiment are up to this value.

The equilibrium position $z$ of the perturbated by the magnetic attraction pendulum with length $L_\mathrm{m} \approx 55\,\mathrm{cm}$ and coil mass $m_\mathrm{m}=1.18\,\mathrm{g},$ is determined by the minimum of the potential energy\cite{LL}
%%%
\bear
&& U_\mathrm{m}(z) = \frac{1}{2} \frac{m_\mathrm{m} g}{L_\mathrm{m}}(y_\mathrm{m} = x-z)^2 - M(z)I^2, 
\qquad \mbox{where:}
\nonumber 
\\
&& M(z)\equiv\frac{2 \pi R_\mathrm{m} N }{I} \left \{ A_\varphi \equiv \frac{\mu_0}{4 \pi} \frac{4NI}{\kappa} \left [ \left ( 1 - \frac{\kappa^2}{2} \right ) \mathrm{K}-\mathrm{E} \right ] \right \}, \nonumber 
\\
&& \kappa=\frac{1}{\sqrt{1+\epsilon^2}}, \qquad \epsilon(z)=\frac{z}{D_\mathrm{m}}, \nonumber 
\\
&&  \mathrm{K}(\kappa^2) \equiv \int_0^{\pi/2} \frac{\mathrm{d}\theta}{\sqrt{ 1-\kappa^2  \sin^2{\theta} }}, \nonumber 
\\
&&  \mathrm{E}(\kappa^2)  \equiv \int_0^{\pi/2} \sqrt{ 1-\kappa^2  \sin^2{\theta} } \, \mathrm{d}\theta,  \nonumber
\eear
%%%
and by the zeroing of the force\cite{LL}
%%%
\bear
&& F_\mathrm{m}(z) = -\mathrm{d}_z U_\mathrm{m} = \frac{m_\mathrm{m} g}{L_\mathrm{m}}(x-z) + \left (  \frac{1}{2} I^2 \mathrm{d}_z M = - 2 \pi R_\mathrm{m} NI B_r(z) \right )=0,
\quad\mbox{where:}
\nonumber 
\\
&& B_r(z)  \equiv  -\mathrm{d}_z  A_\varphi =  \frac{\mu_0}{4 \pi} \frac{NI}{D_\mathrm{m}} \frac{4\epsilon}{\sqrt{1+\epsilon^2}} \left ( -\mathrm{K} + \frac{1+2 \epsilon^2}{2 \epsilon^2} \mathrm{E} \right ),
\nonumber
\\\nonumber
&&
\frac{\mathrm{d}\mathrm{K}}{\kappa\mathrm{d} \kappa}=
2\frac{\mathrm{d}\mathrm{K}}{\mathrm{d} \kappa^2}
=\frac{\mathrm{E}}{\kappa^2(1-\kappa^2)}
-\frac{\mathrm{K}}{\kappa^2},\qquad
\frac{\mathrm{d}\mathrm{E}}{\kappa\mathrm{d} \kappa}=
2\frac{\mathrm{d}\mathrm{E}}{\mathrm{d} \kappa^2}
=\frac{\mathrm{E}-\mathrm{K}}{\kappa^2}.
\eear
%%%
All formulae are given in the most rigorous, logically and sequential way possible, see for example arbitrary encyclopedia 
on theoretical physics.\cite{LL}
The final formulae for the effective potential energy $U_\mathrm{m}$ and the force $F_\mathrm{m}$ are expressed by the mutual inductance $M$,
radial component of the magnetic field $B_r$ and the azimuthal component 
of the vector-potential $A_\varphi$. The force between two coaxial coils is derived in every complete text on electrodynamics.
In most software systems the argument of elliptic integrals 
$\mathrm{E}$ and $\mathrm{K}$ is $\kappa^2$.
The mutual inductance between the coils can be determined experimentally by applying a current through one of the coils and measuring the electromotive voltage $\mathcal{E}_2 =-M \mathrm{d}_t I_1$ of the other.
One of the methods for measuring the mutual inductance between the coils is, for example, 
applying a DC current trough one of the coils,
fast switching off the current and
measuring the peak voltage on the other.
As a rule however, for practical realizations a harmonic AC current is applied and the voltage is measured by a lock-in but those are technical details.
The radial magnetic field $B_r$ is expressed by $z$-derivative of the azimuthal component of the vector potential $A_\varphi$. 
The minus sign in the effective potential energy $-MI^2$ has the same nature\cite{LL} as the minus sign of the effective electric potential energy 
$-C\mathcal{E}^2/2$.

\begin{figure}[t]
\includegraphics[width=18cm]{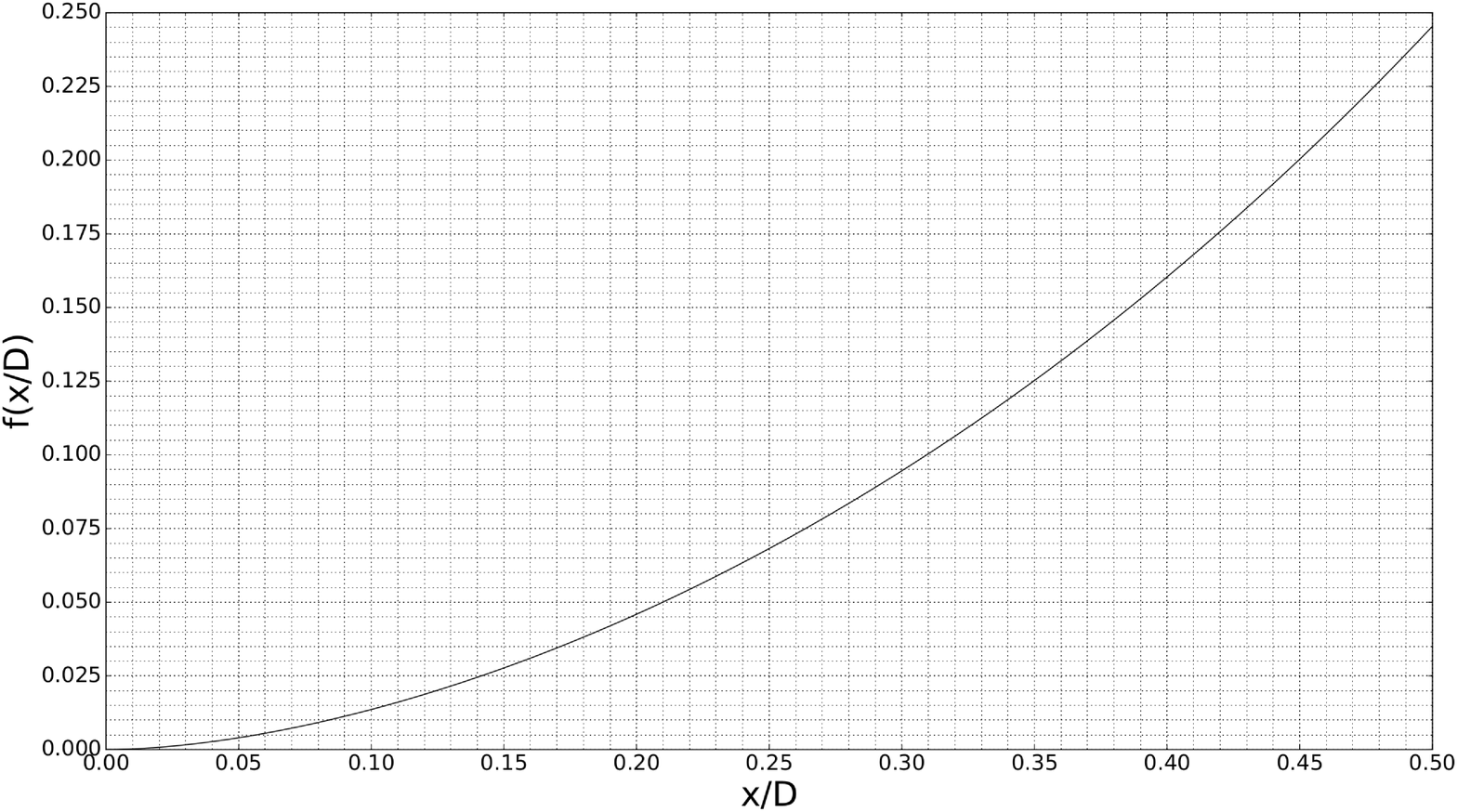}
\caption{The correction function $f_\mathrm{m}(x_\mathrm{m}/D_\mathrm{m})$ as a function of the dimensionless ratio of the equilibrium displacement $x_\mathrm{m}$ and the coils diameter $D_\mathrm{m}$ (index m is omitted for brevity).}
\label{fdelta}
\end{figure}

The experimental method of the magneto-static experiment is slightly different.
Instead of a fixed set of voltages $\mathcal{E}$ and block movement changing $x$, we now fix $x$ and with a voltage source and potentiometer vary the total current $I$ passing successively through both closely separated coils $\epsilon=z/D_\mathrm{m} \ll 1$.
Gradually increasing the current, which is a control parameter now, the small oscillations frequency decreases, and at a definite critical current $I$ the system loses stability and the pendulum coil sticks to the one fixed on the block.
The solution of the magneto-static problem 
%%%
\begin{equation}
F_\mathrm{m}(z_\mathrm{m})=-\mathrm{d}_z U_\mathrm{m}=0  \quad \& \quad k_\mathrm{m}(z_\mathrm{m})=\mathrm{d}_z F_\mathrm{m}=-\mathrm{d}_z^2 U_\mathrm{m}=0
\label{minfl}
\nonumber
\end{equation}
%%%
determines the distance $x_\mathrm{m}$ between the coils at the potential inflection point and gives the condition
%%%
\begin{equation}
\mu_0 I^2 =\mathcal{F}_\mathrm{m}\equiv \frac{m_\mathrm{m} g}{2 L_\mathrm{m} N^2 D_\mathrm{m}} x_\mathrm{m}^2 [1 + f_\mathrm{m}(x_\mathrm{m}/D_\mathrm{m})],
\nonumber
\end{equation}
%%%
where the correction function $f_\mathrm{m}>0$ is depicted in Figure~\ref{fdelta} and tabulated in Table~\ref{tdelta}. 
The experimental data processing is related to fitting of experimental data by the linear regression in the ($\mathcal{F}_\mathrm{m}$, $I^2$)~plane.
The parameters of the magnetic experiment are similar to the electric one, see also the photo of the experimental setup.\cite{EPO4}
Thus the constant $\mu_0$ is determined in an explicit form by the coefficient in the linear regression of the experimental data as $\varepsilon_0$ by the electric experiment.
%%%%%%%
\begin{center}
\begin{table}[t]
\begin{tabular}{| c | c || c | c || c | c || c |  c || c | c |}
		\hline
			\multicolumn{1}{| c |}{ } & \multicolumn{1}{ c ||}{ } & \multicolumn{1}{ c |}{ } & \multicolumn{1}{ c ||}{ } & \multicolumn{1}{ c |}{ } & \multicolumn{1}{ c ||}{ } 
			& \multicolumn{1}{ c |}{ } & \multicolumn{1}{ c ||}{ } & \multicolumn{1}{ c |}{ } & \multicolumn{1}{ c |}{ } \\
			\boldmath$x/D$ & \boldmath$f(x/D)$ & \boldmath$x/D$ &\boldmath$f(x/D)$ & \boldmath$x/D$ & \boldmath$f(x/D)$ & \boldmath$x/D$ & \boldmath$f(x/D)$  & \boldmath$x/D$ & \boldmath$f(x/D)$ \\[15pt] \hline 
			0.005	&	0.0001	&	0.105	&	0.0148	&	0.205	&	0.0480	&	0.305	&	0.0976	&	0.405	&	0.1644	\\	\hline
0.010	&	0.0002	&	0.110	&	0.0162	&	0.210	&	0.0502	&	0.310	&	0.1005	&	0.410	&	0.1682	\\	\hline
0.015	&	0.0005	&	0.115	&	0.0175	&	0.215	&	0.0522	&	0.315	&	0.1034	&	0.415	&	0.1721	\\	\hline
0.020	&	0.0008	&	0.120	&	0.0188	&	0.220	&	0.0545	&	0.320	&	0.1066	&	0.420	&	0.1760	\\	\hline
0.025	&	0.0012	&	0.125	&	0.0202	&	0.225	&	0.0566	&	0.325	&	0.1096	&	0.425	&	0.1800	\\	\hline
0.030	&	0.0016	&	0.130	&	0.0216	&	0.230	&	0.0589	&	0.330	&	0.1126	&	0.430	&	0.1840	\\	\hline
0.035	&	0.0021	&	0.135	&	0.0231	&	0.235	&	0.0613	&	0.335	&	0.1156	&	0.435	&	0.1880	\\	\hline
0.040	&	0.0027	&	0.140	&	0.0245	&	0.240	&	0.0634	&	0.340	&	0.1190	&	0.440	&	0.1920	\\	\hline
0.045	&	0.0034	&	0.145	&	0.0262	&	0.245	&	0.0659	&	0.345	&	0.1221	&	0.445	&	0.1965	\\	\hline
0.050	&	0.0040	&	0.150	&	0.0278	&	0.250	&	0.0684	&	0.350	&	0.1252	&	0.450	&	0.2006	\\	\hline
0.055	&	0.0048	&	0.155	&	0.0294	&	0.255	&	0.0709	&	0.355	&	0.1287	&	0.455	&	0.2047	\\	\hline
0.060	&	0.0056	&	0.160	&	0.0312	&	0.260	&	0.0732	&	0.360	&	0.1319	&	0.460	&	0.2092	\\	\hline
0.065	&	0.0064	&	0.165	&	0.0328	&	0.265	&	0.0757	&	0.365	&	0.1355	&	0.465	&	0.2134	\\	\hline
0.070	&	0.0073	&	0.170	&	0.0345	&	0.270	&	0.0783	&	0.370	&	0.1390	&	0.470	&	0.2181	\\	\hline
0.075	&	0.0083	&	0.175	&	0.0365	&	0.275	&	0.0810	&	0.375	&	0.1423	&	0.475	&	0.2223	\\	\hline
0.080	&	0.0092	&	0.180	&	0.0382	&	0.280	&	0.0837	&	0.380	&	0.1460	&	0.480	&	0.2270	\\	\hline
0.085	&	0.0103	&	0.185	&	0.0401	&	0.285	&	0.0864	&	0.385	&	0.1497	&	0.485	&	0.2317	\\	\hline
0.090	&	0.0114	&	0.190	&	0.0421	&	0.290	&	0.0891	&	0.390	&	0.1530	&	0.490	&	0.2361	\\	\hline
0.095	&	0.0125	&	0.195	&	0.0440	&	0.295	&	0.0919	&	0.395	&	0.1568	&	0.495	&	0.2408	\\	\hline
0.100	&	0.0136	&	0.200	&	0.0461	&	0.300	&	0.0947	&	0.400	&	0.1606	&	0.500	&	0.2456	\\	\hline

		\hline
\end{tabular}
	\caption{The correction function $f_\mathrm{m}(x_\mathrm{m}/D_\mathrm{m})$ tabulated as a function of the dimensionless ratio of the equilibrium displacement $x_\mathrm{m}$ and the coils diameter $D_\mathrm{m}$ (index m is omitted for brevity).}
	\label{tdelta}
\end{table}
\end{center}
%%%%%%%
%%%%%%%
%%%%%%%
We can use linear regression or simply divide the measured current by the ammeter from the right side.
At small values of the parameter $\delta_\mathrm{m}=x_\mathrm{m}/D_\mathrm{m}$ when the coils are separated at a distance much less than their diameter, the approximate formulas for elliptic integrals\cite{JEL}
%%%
\bear
&& \mathrm{K} = \Lambda + \frac{1}{4} (\Lambda-1){\kappa^\prime}^2 + \frac{9}{64} \left ( \Lambda - \frac{7}{6} \right ) {\kappa^\prime}^4 
+\frac{25}{256}\left(\Lambda - \frac{37}{30}\right){\kappa^\prime}^6
+ \cdots, \nonumber 
\\
&& \mathrm{E} = 
1+\frac{1}{2} \left ( \Lambda - \frac{1}{2} \right ) {\kappa^\prime}^2 
+\frac{3}{16}\left(\Lambda - \frac{13}{12}\right){\kappa^\prime}^4
+\frac{15}{128}\left(\Lambda - \frac{6}{5}\right){\kappa^\prime}^6
+ \cdots, \nonumber 
\\
&&\kappa^\prime=\sqrt{1-\kappa^2}=4\mathrm{e}^{-\Lambda}
=\frac{\epsilon}{\sqrt{1+\epsilon^2}}
\ll1, \nonumber
\\
&&
\Lambda=\ln{\left (\frac{4}{\kappa^\prime} \right )}
=-\ln\epsilon+\frac12\ln(1+\epsilon^2)+\ln4
\nonumber
\eear
%%%
give
%%%
\begin{equation}
f_\mathrm{m}(\delta_\mathrm{m}) \approx \frac{1}{16} \left ( -5 + 6\ln{\frac{8}{\delta_\mathrm{m}}} \right ) \delta_\mathrm{m}^2 + \mathcal{O}(\delta_\mathrm{m}^4)
\label{fcorr}
\nonumber
\end{equation}
%%%
and this approximation for the system parameters $m_\mathrm{m}$ and $L_\mathrm{m}$ provides percent accuracy in the $\mu_0$ determination. 
The well-known from the mathematical analysis $\mathcal O$ function means which power of the argument is neglected in the current approximation. 
%%%%%%%%%%%%%%%%%%

\subsection{Determination of the light velocity}

At known $\varepsilon_0$ and $\mu_0$, the velocity of light  $c=1/\sqrt{\varepsilon_0 \mu_0}$ is also determined with percent accuracy and this also is the accuracy of the standard multimeter for current $I$ and voltage $\mathcal{E}$ measurements, as is the precision of the measured distances with 0.5~mm accuracy also.
The largest error is in the distance $x$ measurement and it may be useful a magnifying glass to be added to the experimental setup after power source shut off.
Both methods have in common the lack of forces measurement with an analytical (electronic) scale, which significantly lowers the price of the experimental setup and makes it suitable for popularization even for 10-ager terminators of expensive equipment.
Both electrostatic and magneto-static experiments have also in common the usage of the catastrophe theory that we mention in the next section.

%%%%%%%%%%%%%%%%%%%%%%%%%%%%%%%%%%
%
%%% Catastrophe theory					
%
%%%%%%%%%%%%%%%%%%%%%%%%%%%%%%%%%%
\section{Catastrophe theory}

The current for the magneto-static experiment can be fixed too, and thus both experiments control parameter will be the distance $x$ between the coils or the capacitor plates after the voltage disconnect and short-circuiting.
It is convenient to consider $x$ as a parameter of the total potential energy $U(z,x)$ dependence on the distance $z$ between the plates or coils at a switched circuit.
Around the minimum we have:
%%%
\begin{equation}
U(z) \approx \frac{1}{2} k(z,x)(z-z_0)^2.
\nonumber
\end{equation}
%%%
The position of the minimum $z_0$ is determined by the zeroing of the force
%%%
\begin{equation}
F(z_0)=\left. -\mathrm{d}_z U(z) \right |_{z=0}=0
\nonumber
\end{equation}
%%%
and the small oscillations frequency is determined by the second derivative in the minimum
%%%
\begin{equation}
\omega = \sqrt{k(z_0,x)/m}, \qquad \left. k(z_0(x))=\mathrm{d}_z^2U(z,x) \right|_{z=z_0} \ge 0.
\nonumber
\end{equation}
%%%
Let us look at the second derivative $k(z_0(x))$ behaviour, i.e. the system rigidity around this minimum when the control parameter $x$ is varied.
Gradually decreasing the parameter $x,$ at some critical value $x_\mathrm{c}$ the second derivative in the minimum becomes zero $k(z_0, x_\mathrm{c})=0$.
If we analyse the potential energy as a function of two variables $U(z,x)$, we have the mathematical problem for finding the solution $(z_\mathrm{c},x_\mathrm{c})$ of the system
%%%
\begin{equation}
\partial_z U(z,x)=0 \quad \& \quad \partial_z^2 U(z,x)=0.
\nonumber
\end{equation}
%%%
At close proximity to the thus determined values of the potential energy variables we have the approximation
%%%
\bear
&& U = U_\mathrm{c} + U_{zx}(z-z_\mathrm{c})(x-x_\mathrm{c}) + \frac{1}{3!} U_{zzz}(z-z_\mathrm{c})^3, \nonumber 
\\
&& U_{zx} = \partial_z \partial_x U(z_\mathrm{c},x_\mathrm{c}) < 0, \qquad  U_{zzz}=\partial_z^3U(z_\mathrm{c},x_\mathrm{c})>0, \qquad U_\mathrm{c} = U(z_\mathrm{c},x_\mathrm{c}). 
\nonumber
\eear
%%%
Let us introduce new variables
%%%
\begin{equation}
\overline{x}=z-z_\mathrm{c}, \qquad \overline{b}=-\frac{3! (U-U_\mathrm{c})}{U_{zzz}}, \qquad \overline{a}=\frac{3!U_{zx}}{U_{zzz}}(x-x_\mathrm{c}),
\nonumber
\end{equation}
%%%
then the potential energy approximation has the standard form of the canonical fold catastrophe from the catastrophe theory
%%%
\begin{equation}
\overline{a}\,\overline{x}+\overline{x}^3+\overline{b}
=\frac{\mathrm{d}}{\mathrm{d}\overline{x}}
\left(\frac12\overline{a}\,\overline{x}^2+\frac14\overline{x}^4+\overline{b}\overline{x}\right)=0.
\nonumber
\end{equation}
%%%
This fold in the space $(\overline{a},\overline{b},\overline{x})$ is presented 40~times and the corresponding formula 12~times in the well-known reference monograph on catastrophe theory and its applications by Tim~Poston and Ian~Stewart.\cite{PostStew}
Let us review the used terminology.

The pendulum transition, which at a critical value of the current $I$, voltage $\mathcal{E}$ or the distance from equilibrium $x$ suddenly rushes towards the block is an example for the so called catastrophic jumps by Ren\'e Thom\cite{Thom} and Cristopher Zeeman.\cite{{ZeemanCM}}
The variables $x$, $\mathcal{E}$ or $I$ are called control variables (or control parameters) and $z$ is called a behaviour variable (or state variable). The catastrophic jumps occur when smooth variations of controls cause a discontinuous change of state.
In other words, the variable $x$ is a control parameter and the distance $z$ between the coils or capacitor plates is a behaviour variable.
The variable $z$ has a catastrophic change when a smooth variation of $x$ takes place around the critical value $x_\mathrm{c}$.
Without referring the catastrophe theory notions explicitly, such behaviour can be found in many physical problems: 
stability of orbits in the field of a black hole (briefly mentioned below), 
appearance of p-, d-,  f-, and g-electrons in atoms with different Z,
critical point, corresponding states rule and Landau theory of second order phase transitions,  
plane flow of compressional gases, see the well-known Landau and Lifshitz encyclopaedia.\cite{LL}
And there are applications in such fields like heartbeat and propagation of a nerve impulse.\cite{Zeeman,ODE}
Landau concepts of description of phase transition by breaking symmetry order parameter replaced science of type of zoology in a unified theory.\cite{PatashinskiPokrovsky}
It is interesting that even biological phenomena can be described by differential equations similar to the kinetics of the order parameter.\cite{LL}

Our experimental setup is to a large extent influenced by the Zeeman catastrophic machine\cite{ZeemanCM} and by Tim~Poston's work on Do-it-yourself catastrophe machine.\cite{Poston}
In our machine the rubber elastics are replaced
by force lines of the electric and magnetic field.
In the same intuitive manner in which 
Faraday introduced force lines and concepts of a field in the mathematical Physics.
Do-it statement does not refer to funding restrictions, we introduce a new idea 
for the usage of the notions of the catastrophe theory in
the methodology of a student laboratory.
Concerning the high school students, they are potential terminators of precise scales. 

\begin{figure}[h]
\includegraphics[width=18cm]{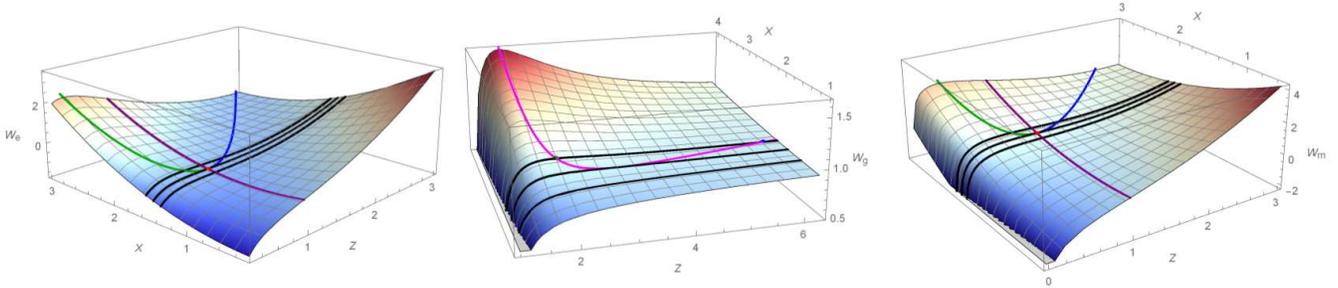}
\caption{
The potential surfaces $W_\mathrm{e}$ (left), $W_\mathrm{g}$ (center) and $W_\mathrm{m}$ (right). 
The right branch of the curves (blue in the colour version) lines represent the stable local minima of the potential energy as function of $Z$ at fixed values of $X$.
The left branch of the curves (green in the colour version) show the local unstable maxima.
Those two branches (the green and blue lines) join at the critical point (red in the colour version) at which minima and maxima annihilate for the critical value $X_\mathrm{c}.$
For the gravitational problem local extrema are depicted by magenta in the colour version.
The 3 parallel black curves over all 3 surfaces are copied in the next Fig.~\ref{curves}.
Those curves demonstrate local extrema for $X>X_\mathrm{c},$
monotonous dependence for $X<X_\mathrm{c},$
and most important 
the catastrophic behavior at the critical value $X=X_\mathrm{c}.$
         }
\label{surfaces}
\end{figure}

\begin{figure}[h]
\includegraphics[width=18cm]{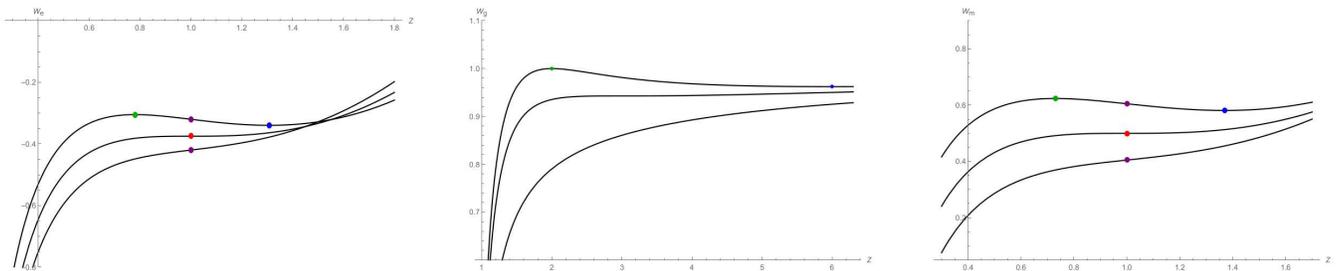}
\caption{Three sections of the potential surfaces  close to the critical point for electric $W_\mathrm{e}$ (left), gravitational $W_\mathrm{g}$ (center) and magnetic $W_\mathrm{m}$ (right) problem.
The points of the unstable maxima are marked in green in the colour version, while 
the points of local minima are marked by blue
for $X>X_\mathrm{c}.$
For the critical value $X=X_\mathrm{c}$
maximums and minimums merge in an inflection point marked by red in the colour version. 
For $X<X_\mathrm{c}$ the potential curves are monotonous without local extrema.
The zeros of the second derivative between the minumum and the maximum of the potential curves are shown in purple in the colour version.}
\label{curves}
\end{figure}

The theory of the described experiment is related to analysis of the potential surfaces derived in the appendices for the electric $W_\mathrm{e}$ and magnetic $W_\mathrm{m}$ problem. 
In Fig.~\ref{surfaces} the surfaces are depicted in dimensionless
variables
\bear
&&
\nonumber
W_\mathrm{e}(Z,X)=\frac12 (X-Z)^2 - \frac{1}{2Z}
,\\
&&
\nonumber
W_\mathrm{g}(Z,X)
=\sqrt{\left(1-\frac1{Z}\right)\left(1+\frac{X^2}{Z^2}\right)}
,\\
&&
\nonumber
W_\mathrm{m}(Z,X)=\frac12 (X-Z)^2 +\ln Z.
\eear
For the gravitational problem of stability of a circular orbit around a black hole\cite{LL} $W_\mathrm{g}=U/mc^2,$ $Z=r/r_g$, $X=M/mcr_g$ is the dimensionless angular momentum and $r_g$ is the Schwarzschild radius.
We refer to black holes because ``collapse'' of the plates of the capacitor or the
joining of the coils of the magnetic pendulum is analogous to the recently observed merging
of black holes,\cite{Abbott} 
which approximately can be described using fold instability.
The sections  $W(Z,X)$ in Fig.~\ref{curves} are given for 
3 typical sections  values 
$W(Z;X)$ for $X<X_\mathrm{c}$, $X=X_\mathrm{c}$ and $X>X_\mathrm{c}$.
 %%%%%%%
%%%%%%%
%%%%%%%
%%%%%%%
%%%%%%%
\begin{figure}[h]
\includegraphics{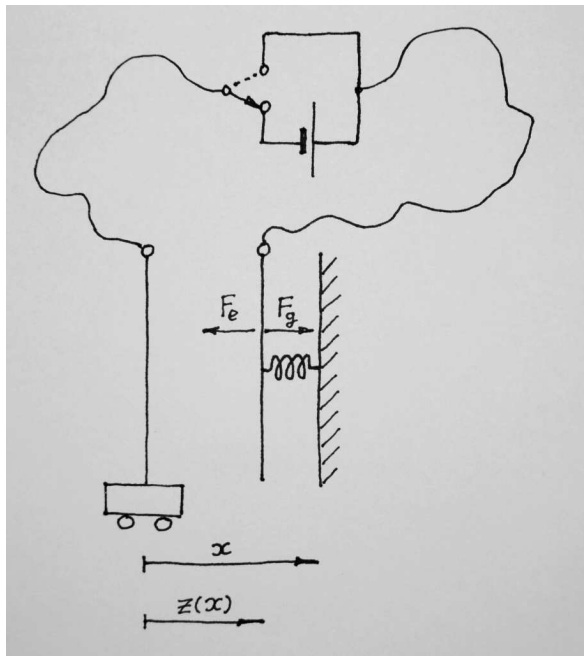}
\caption{Scheme of the method for electrostatic measurement: a picture of the experimental setup is given in the EPO4 problem.
Under the action of the electric field created by the batteries, the movable plate of the capacitor deforms the spring with effective rigidity $\varkappa=m_\mathrm{e}g/L_\mathrm{e}.$
In equilibrium position $z(x)$ ``the elastic force'' 
$F_\mathrm{g}=\varkappa(x-z)$ is compensated by the electric force $F_\mathrm{e}=-\frac12 \varepsilon_0 (E=\mathcal{E}/z)^2(S=\pi R_\mathrm{e}^2).$ 
We prefer an expression in which one can easily trace the origin of the different multipliers. 
We gradually decrease $x$ and at some critical value
$x_\mathrm{c}$ the equilibrium position
$z_\mathrm{c} \approx x_\mathrm{c}/3$ loses stability and a catastrophe happens.
The pendulum (the suspended plate of the capacitor) suddenly sticks to the fixed one at $z=0.$ When the switch is changed to upper position, the pendulum minimizes its gravitational energy 
$\varkappa (x_\mathrm{c}-z)^2/2$ 
only and $z=x_\mathrm{c}.$ In a good approximation
$x_\mathrm{c}^3 \propto \mathcal{E}^2.$
In the magneto-static experiment the plates are substituted with coils with parallel currents and 
$x_\mathrm{c}^2 \propto I^2.$
Similar catastrophe happens with a circular orbit around a black hole too.}
\label{scheme}
\end{figure}
%%%%%%%

The St.~Clement of Ohrid University students use a catastrophic machine for measurement of the fundamental constant velocity of light. We review some technical details in the next section.

%%%%%%%%%%%%%%%%%%%%%%%%%%%%%%%%%%
%
%%% Technical details
%
%%%%%%%%%%%%%%%%%%%%%%%%%%%%%%%%%%
\section{Technical details}

In numerical analysis of the problem when the total potential energy for both electrostatic and magneto-static problems are programmed as functions $U(z,x)$, the inflection point is found via a solution of the corresponding system for zeroing of the force $F$ and the rigidity of the system $k$.
With thus found current $I$ or voltage $\mathcal{E}$ critical values, the universal scaling functions $\delta=x/D$ can be determined by the numerical solution
%%%
\bear
&& f_\mathrm{e}(\delta_\mathrm{e}=x_\mathrm{e}/D_\mathrm{e} )  = 1 - \frac{27 \pi L_\mathrm{e} D_\mathrm{e}^2 \varepsilon_0 \mathcal{E}^2}{32 m_\mathrm{e} g x_\mathrm{e}^3}, \nonumber 
\\
&& f_\mathrm{m}(\delta_\mathrm{m}=x_\mathrm{m}/D_\mathrm{m} ) = \frac{2 L_\mathrm{m} N^2 D_\mathrm{m}}{\mu_0 I^2 m_\mathrm{m} g x_\mathrm{m}^2}-1. \nonumber
\eear
%%%
There are analytical methods, of course, that give power series and the first correction was given for homework to the participated students in EPO4 with a Sommerfeld prize of 137~DM.
And the high school students had to derive the main term with $f_\mathrm{e} \approx 0$ and $f_\mathrm{m} \approx 0$ during the Olympiad too.\cite{EPO4}

%%%%%%%%%%%%%%%%%%%%%%%%%%%%%%%%%%
%
%%% History notes
%
%%%%%%%%%%%%%%%%%%%%%%%%%%%%%%%%%%
\section{History notes}

Immediately after realizing that there is current in the magnetic field equation $\textbf{j}+\varepsilon_0 \partial_t \textbf{E}$, Maxwell understood that the velocity of light can be determined from purely static separate measurements of electric and magnetic forces connected with the electromagnetic stress tensor and its energy density 
$\frac12 \varepsilon_0 E^2+\frac{1}{2}B^2/\mu_0$.\cite{Maxwell} 
If the product of unit current $\mathrm{e}_{_I}$ and unit voltage $\mathrm{e}_\mathcal{_E}$ gives the mechanical unit for power, then $c=1/\sqrt{\varepsilon_0 \mu_0}$ in any choice of units.
The $ \varepsilon_0$ and $\mu_0$ numerical values is a matter of choice and convenience, for instance in Gaussian units $4 \pi \varepsilon_0=1$ and $\mu_0/4 \pi=1$ and these relict multipliers participate in our formulas.
In Lorentz-Heaviside units $\varepsilon_0=1$ and $\mu_0=1$ and naturally $c=1$.
This is practice in the modern metrology, the velocity of light is not measured from a long time and the convention $c=299~792~458$~m/s is used.~\cite{Mohr} The unit meter is redefined at a fixed time standard.
The same can be said for the Ampere, the unit of current is fixed in 1948 in order $\mu_0=4\pi\times10^{-7}\;\mathrm{NA^{-2}}$.
In this sense ``measurement the speed of light'' only marks an important stage of the development of physics. 
In common language using, for example Google, the ``speed of light'' is almost twenty times more frequently used than ``velocity of light'' but in science in the titles of the arXiv
e-prints the frequencies are comparable.
But even now, when the student measure the mechanical force of the electric field the tutorial\cite{MIT} says:
\textit{Congratulations you have just measured one of the fundamental constants of nature!}
For $\mu_0$ again\cite{MIT} with one extra comma:
\textit{Congratulations, you have just measured one of the fundamental constants of nature!}. As in biology, the individual development repeats the evolutionary one. That is why we are saying to the students that they ``measure'' fundamental constants, not: \textit{Congratulations your multimeter is still OK!}.

The purpose of our methodical experiment is to guide the students through the development of the electrodynamics using for fun a catastrophic machine that can be built in a day, costs \pounds20 and has a percent accuracy in case of precise work.
But we use catastrophe machine not by funding restrictions but to demonstrate how an good mathematical idea\cite{Poston} can be used in student laboratory experiment.

Organizing of a Olympiad with 137 participants and giving the setup to every one we had no possibility to buy for everybody electronic scale. That is why we decided to apply catastrophic theory which requires to measure only distances but not forces.
Of course, for students labs the usage of measurement of forces or balance of scales is a tradition coming from the time of Maxwell.\cite{MaxwellExp} Let us mention the setup of
Berkeley university,\cite{Purcell}
MIT,\cite{MIT}
University of Sofia,\cite{Gourev}
and the Gymnasium in Breziche.\cite{Breziche}

%%%%%%%%%%%%%%%%%%%%%%%%%%%%%%%%%%
%
%%% Conclusion
%
%%%%%%%%%%%%%%%%%%%%%%%%%%%%%%%%%%
\section{Conclusions}

This experimental setup is a part of the Physics faculty of St.~Clement of Ohrid University program for development of cheap experimental setups for fundamental constants measurements, see for example the description of the setup for measurement of Planck constant by electrons\cite{PlanckLandauer}
and the measurement of speed of light by analytical scales.\cite{Gourev}
The experimental setups can be constructed even in high (secondary) school laboratories and the corresponding measurements can be conducted by the high (secondary) school students.
The authors are grateful to 137~participants (students and teachers) in EPO4 where the described experimental setup in this article was used and a dozen students measured $c$ and derived the formulas.~\cite{EPO4}

In general, we can conclude that notions of catastrophe theory can be very useful 
for invention of new set-ups in student laboratory of physics. 
This is a style of thinking in a broad problems in science and technology. 

%%%%%%%%%%%%%%%%%%%%%%%%%%%%%%%%%%
%
%%% Acknowledgements
%
%%%%%%%%%%%%%%%%%%%%%%%%%%%%%%%%%%
\acknowledgments{}

The Olympiad was held with the cooperation of Faculty of Physics of St.~Clement of Ohrid University at Sofia
-- special gratitude to the dean prof.~A.~Dreischuh and
also to
president of Macedonian physical society assoc.~prof.~B.~Mitrevski, 
and the president of the Balkan Physical Union acad.~A.~Petrov.

We, the EPO4 organizers, are grateful to the high school participants who managed to derive the  $\varepsilon_0 \mathcal{E}^2$ and $\mu_0 I^2$ formulas without the correction functions $f_\mathrm{e}$ and $f_\mathrm{m}.$

The authors,  including the EPO4 champion Dejan Maksimovski  who measured the velocity of light with 1~\% accuracy, are thankful to the university students from Skopje Biljana Mitreska and Ljupcho Petrov, who during the night after the experimental part of the Olympiad, solved a significant part of the derived here correction functions $f_\mathrm{e}$ and $f_\mathrm{m}$ used for the accurate determination of $\varepsilon_0$, $\mu_0$ and $c$ by the used electrostatic and magneto-static experiments.

%%%%%%%%%%%%%%%%%%%%%%%%%%%%%%%%%%
%
%%% Bibliography
%
%%%%%%%%%%%%%%%%%%%%%%%%%%%%%%%%%%

%%%%%%%%%%%%%%%%%%%%%%%%%%%%%%%%%%

%%%%%%%%%%%%%%%%%%%%%%%%%%%%%%%%%%
%
%%% Appendix
%
%%%%%%%%%%%%%%%%%%%%%%%%%%%%%%%%%%
\appendix
\section{Effects of ends correction}
%%%%%%%%%%%%%%%%%%

Let us look in details at the inflection point of the electrostatic experiment
%%%
\begin{equation}
U_\mathrm{e}(z,x_\mathrm{e}) = \frac12 \varkappa (x-z)^2 - \frac12 C(z)\mathcal{E}^2, \qquad
 \varkappa=\frac{m_\mathrm{e}g}{L_\mathrm{e}}.
\nonumber
\end{equation}
%%%
For alleviation of the further notation let us introduce the quantity with length dimension
%%%
\begin{equation}
a^3 \equiv \frac{\varepsilon_0 S \mathcal{E}^2}{\varkappa},
\nonumber
\end{equation}
%%%
the dimensionless lengths
%%%
\begin{equation}
X = \frac{x}{a}, \qquad
Y= \frac{y}{a}, \qquad
Z= \frac{z}{a}
\nonumber
\end{equation}
%%%
and the small dimensionless parameter
%%%
\begin{equation}
\eta \equiv \frac{a}{\pi R_\mathrm{e}} \ll 1.
\nonumber
\end{equation}
%%%
The potential energy in these variables takes the form
%%%
\begin{equation}
W(Z,X) \equiv \frac{U_\mathrm{e}(z,x_\mathrm{e})}{\varkappa a^2} = \frac12 (X-Z)^2 - \frac{1}{2Z} + \frac12 \eta \ln{Z} - \frac12 \overline{C}_\mathrm{H} - \frac12 \eta \ln\frac{\eta}{\sqrt{\pi}},
\nonumber
\end{equation}
%%%
or for negligible $\eta$
%%%
\begin{equation}
W(Z,X) \equiv \frac{U_\mathrm{e}(z,x_\mathrm{e})}{\varkappa a^2} = \frac12 (X-Z)^2 - \frac{1}{2Z}.
\nonumber
\end{equation}

The equation for zeroing of the second derivative at the inflection point 
%%%
\begin{equation}
k_\mathrm{e}(z) = \mathrm{d}^2_z  U_\mathrm{e} = \varkappa \mathrm{d}^2_{_Z} W = \left ( 1 -  \frac{1}{Z^3} - \frac{\eta}{2Z^2} \right ) \varkappa = 0
\nonumber
\end{equation}
%%%
we solve with the method of successive approximation in the series of $\eta$ as further we take into account only the linear correction and omit the negligible  for our experiment $\eta^2$ and higher powers.
In the zeroth approximation we have $Z^{(0)}=1$ and in the first one multiplying with $Z^3$ we have
%%%
\begin{equation}
Z^3 = 1 + \frac{\eta}{2} Z^{(0)} \approx 1 + \frac{\eta}{2}, \qquad
Z \approx 1 + \frac{\eta}{6}.
\nonumber
\end{equation}
%%%
With the thus determined dimensionless distance between the plates, we derive the pendulum deviation $Y=X-Z$ from the condition for the balance of the forces
%%%
\begin{equation}
F = -\mathrm{d}_z  U_\mathrm{e} = -\varkappa a \left [  \mathrm{d}_{_Z}  W=-(X-Z) + \frac12 \left (\frac{1}{Z^2} + \frac{\eta}{Z} \right ) \right ]=0,
\nonumber
\end{equation}
%%%
which gives
%%%
\begin{equation}
X=\frac32 + \frac{\eta}{2}, \qquad
Y=\frac12 + \frac{\eta}{3}, \qquad
X^3 = \frac{27}{8} (1+\eta),
\nonumber
\end{equation}
%%%
using $(1+\eta)^n \approx 1 + n\eta.$ In the zeroth approximation we have
%%%
\begin{equation}
z^{(0)} = a = \frac23 x^{(0)}, \qquad
\eta = \frac{a}{\pi R_\mathrm{e}} 
=\frac23 \frac{x^{(0)}}{\pi R_\mathrm{e}}\ll1
\nonumber
\end{equation}
%%%
thus
%%%
\begin{equation}
X^3 = \frac{x_\mathrm{e}^3}{a^3} \approx \frac{27}{8} \left (1+\frac{2}{3\pi} \frac{x_\mathrm{e}}{R_\mathrm{e}} \right )
\nonumber
\end{equation}
%%%
and after substitution of $a$ and $D_\mathrm{e}$ we get the formula for $\varepsilon_0 \mathcal{E}^2.$

%%%%%%%%%%%%%%%%%%%%%%%%
\section{Magneto-static pendulum stability analysis}

The magnetic force can be presented by the derivative of the effective potential energy 
%%%
\begin{equation}
U_{_I}(z)=-M(z)I^2.
\nonumber
\end{equation}
%%%
The mutual inductance $M$
describes the magnetic flux one of the coils creates through the other one and the electromotive forces 
%%%
\begin{equation}
\Phi_1=M(z)I_2=\int{\mathbf{B}_2(\mathbf{r}) \cdot \mathrm{d} \mathbf{S}_1}=
\oint{ \mathbf{A}_2 \cdot \mathrm{d} \mathbf{r} }=2\pi R_{\mathrm{m}}A_{\varphi},
\qquad \mathcal{E}_1= M \mathrm{d}_t I_2,
\qquad
\mathcal{E}_2= M \mathrm{d}_t I_1,
\nonumber
\end{equation}
%%%
For equal currents $I_1=I_2=I$ we obtain
for the magnetic force of the parallel attracting currents 
%%%
\begin{equation}
F_{_I}=
-\mathrm{d}_z U_{_{I}}(z) 
= (\varepsilon_{z \varphi \rho}=-1) \, 2 \pi R_\mathrm{m} NI 
   \left ( B_r = - \partial_z A_{\varphi} 
= \frac{\mu_0}{4 \pi} \frac{2NI}{z}(1-f_\mathrm{m}) \right ).
\nonumber
\end{equation}
%%%
The Levi-Civita symbol $\varepsilon_{z \varphi \rho}=-1$ comes from the vector product sign from the Lorentz force $q_e \mathbf{v} \times \mathbf{B}$ acting on every electron flowing through the loop.
The radial magnetic field $B_r$ we define as a product of the elementary formula for the magnetic field of an infinite current $\mu_0(NI)/2 \pi z$ and of the correction multiplier $(1-f_\mathrm{m})$, and for closely separated coils $z \ll D_\mathrm{m}=R_\mathrm{m}/2$ the correction is small $f_\mathrm{m} \ll 1.$ Introducing
%%%
\begin{equation}
b^2 \equiv \frac{\mu_0 R_\mathrm{m} (NI)^2}{\varkappa},
\nonumber
\end{equation}
%%%
the total force
%%%
\begin{equation}
F_\mathrm{m}=-\mathrm{d}_z [ U_{_I}(z) +\frac12 \varkappa(x-z)^2 =U_\mathrm{m}(z)], \qquad
\varkappa=\frac{m_\mathrm{m}g}{L_\mathrm{m}},
\nonumber
\end{equation}
%%%
can be written with the dimensionless variables
%%%
\begin{equation}
X = \frac{x}{b}, \qquad
Y= \frac{y=(x-z)}{b}=X-Z, \qquad
Z= \frac{z}{b}
\nonumber
\end{equation}
%%%
and the condition for the balance of the forces takes the form 
%%%
\begin{equation}
F_\mathrm{m}= \left ( - (X-Z) + \frac{1-f_\mathrm{m}}{Z} \right ) \varkappa b=0.
\nonumber
\end{equation}
%%%
The stability loss at the potential energy minimum is described by the equation
%%%
\begin{equation}
k(z)=\mathrm{d}_z^2 U_\mathrm{m}(z) = \left ( 1 - \frac{1-f_\mathrm{m}}{Z^2}+Z\mathrm{d}_{_Z}f_\mathrm{m} \right )=0.
\nonumber
\end{equation}
%%%
The solution of this equation with successive approximations
%%%
\begin{equation}
\left(Z^{(n+1)}\right)^2=(1-f_\mathrm{m})+\left(Z^{(n)}\right)^3 \mathrm{d}_{_Z} f_\mathrm{m}(Z^{(n)})
\nonumber
\end{equation}
%%%
gives
%%%
\begin{equation}
Z^{(0)}=1, \qquad Z^{(1)}=1-\frac12 f_\mathrm{m}(1) + \frac12 f_\mathrm{m}^\prime(1).
\nonumber
\end{equation}
%%%
The substitution of this solution into the equation for the zeroing of the force gives
%%%
\begin{equation}
X=Z+\frac{1-f_\mathrm{m}}{Z}=2-f_\mathrm{m}+\mathcal{O}(f_\mathrm{m}^2)
\nonumber
\end{equation}
%%%
or
%%%
\begin{equation}
X^2 = \frac{x^2}{b^2}=4(1-f_\mathrm{m})+\mathcal{O}(f_\mathrm{m}^2).
\nonumber
\end{equation}
%%%
For the zeroth approximation, that was derived by high school students, we get
%%%
\begin{equation}
Z^{(0)}= \frac{z^{(0)}}{b}=1,  \qquad
X^{(0)} = \frac{x^{(0)}}{b}=2, \qquad
Y^{(0)}= \frac{y^{(0)}=x^{(0)}-z^{(0)} }{b}=1.
\nonumber
\end{equation}
%%%
The exact formula for the magnetic field obtained by the integration of the Biot–Savart law 
%%%
\begin{equation}
B_0 = \frac{\mu_0}{4 \pi } \frac{2NI}{z}(1-f_\mathrm{m})=\frac{\mu_0}{4 \pi } \frac{2NI}{z} \cdot  \frac{2\epsilon^2}{\sqrt{1+\epsilon^2}} \left ( -\mathrm{K} + \frac{1+2 \epsilon^2}{2 \epsilon^2} \mathrm{E} \right )
\nonumber
\end{equation}
%%%
at closely situated coils
%%%
\begin{equation}
\epsilon=\frac{z^{(0)}}{D_\mathrm{m}}=\frac12 \left ( \delta_\mathrm{m} = \frac{x^{(0)}}{D_\mathrm{m}} \right )
\nonumber
\end{equation}
%%%
after some algebra with elliptic integrals and neglecting terms with higher powers than $\epsilon^2$
%%%
\begin{equation}
\mathrm{K} \approx \Lambda + \frac{1}{4} (\Lambda-1) \epsilon^2, \qquad
\mathrm{E} \approx 1+\frac{1}{2} \left ( \Lambda - \frac{1}{2} \right ) \epsilon^2, \qquad 
\Lambda=\ln{\left (\frac{4}{\epsilon} \right )}
\nonumber
\end{equation}
%%%
we obtain for the correction
%%%
\begin{equation}
f_\mathrm{m} = \left ( -\frac54 + \frac32 \ln{\frac{4}{\epsilon}} \right ) \epsilon^2 =
\frac{1}{16} \left ( -5 + 6\ln{\frac{8}{\delta_\mathrm{m}}} \right ) \delta_\mathrm{m}^2 + \mathcal{O}(\delta_\mathrm{m}^4)
\nonumber
\end{equation}
%%%
and substituting in the formula for $x^2/b^2$ after expressing the definition for $b^2$, gives the used in the experiment formula for $\mu_0 I^2$.

Let us analyse the exact solution too.
The catastrophe fold of the potential should be examined
%%%
\begin{equation}
U_\mathrm{m}(z,x) = \frac12 (x-z)^2 - M(z) I^2, \qquad
\mathrm{d}_z^2 U_\mathrm{m}(z)=0, \qquad
\mathrm{d}_z U_\mathrm{m}(z,x)=0,
\nonumber
\end{equation}
%%%
which in dimensionless variables is
%%%
\begin{equation}
W(Z,X) = \frac{U_\mathrm{m}}{\varkappa b^2} 
= \frac12 (X-Z)^2-\frac{2}{\kappa} 
\left [ \left ( 1 - \frac{\kappa^2}{2} \right )\mathrm{K}-\mathrm{E} \right ], \qquad
\mathcal{D}_\mathrm{m}=\frac{D_\mathrm{m}}{b},
\nonumber
\end{equation}
%%%
or 
%%%
\begin{equation}
W(Z,X) 
= \frac12 (X-Z)^2
+\ln Z -\ln \mathcal{D}_\mathrm{m}
+2\ln\frac{\mathrm{e}}{2}, \qquad 
\mbox{for  } \mathcal{D}_\mathrm{m}\gg1.
\nonumber
\end{equation}
%%%
The critical value $Z_\mathrm{m}$ and $\epsilon_\mathrm{m}=Z_\mathrm{m}/\mathcal{D}_\mathrm{m}$ is determined from the solution of the equation
%%%
\begin{equation}
Z^2 = \frac{1}{(1+\epsilon^2)^{3/2}} [ (1 - \epsilon^2) \mathrm{E} + \epsilon^2 \mathrm{K} ]
\nonumber
\end{equation}
%%%
derived from $\mathrm{d}_z^2 U_\mathrm{m}(z_\mathrm{m})=0$.
For the critical value of $X_\mathrm{m}$, derived from the equation $\mathrm{d}_z U_\mathrm{m}(z_\mathrm{m},x_\mathrm{m})=0$ the software product Mathematica found 

%%%
\bear
&& \delta_\mathrm{m} = \frac{X_\mathrm{m}}{\mathcal{D}_\mathrm{m}} = 2 \epsilon \frac{ (1+\epsilon^2+\epsilon^4)\mathrm{E} - \epsilon^2 (1/2+\epsilon^2)\mathrm{K} }{(1-\epsilon^2)\mathrm{E} + \epsilon^2 \mathrm{K}}, \nonumber \\
&& \frac{X_\mathrm{m}}{Z_\mathrm{m}} = 2 \frac{ (1+\epsilon^2+\epsilon^4)\mathrm{E} - \epsilon^2 (1/2+\epsilon^2)\mathrm{K} }{(1-\epsilon^2)\mathrm{E} + \epsilon^2 \mathrm{K}}. \nonumber
\eear
%%%
Finally, for the correction function defined as
%%%
\begin{equation}
X_\mathrm{m}^2 (1+f_\mathrm{m}) = 4 \quad \mathrm{or} \quad f_\mathrm{m}=\frac{4}{X_\mathrm{m}^2}-1
\nonumber
\end{equation}
%%%
we obtain
%%%
\begin{equation}
1+f(\epsilon)=\frac{ 4(1+\epsilon^2)^{3/2} [(1-\epsilon^2)\mathrm{E} + \epsilon^2 \mathrm{K} }{[2(1+\epsilon^2+\epsilon^4)\mathrm{E} - \epsilon^2 (1/2+\epsilon^2)\mathrm{K}]^2}.
\nonumber
\end{equation}
%%%
This function $f(\epsilon_\mathrm{m})$ together with the relation $\delta(\epsilon_\mathrm{m})$ parametrically determine the function $f(\delta)$ shown in Fig.1~\ref{fdelta}.
And the equation for $\mu_0 I^2$ used for the processing of the experimental data is obtained by expressing 
$x_\mathrm{m}^2 = b^2 X_\mathrm{m}^2$ 
and 
$\delta=x_\mathrm{m}/D_\mathrm{m}
 =X_\mathrm{m}/\mathcal{D}_\mathrm{m}.$
%%%%%%%%%%%%%%%%%%

\end{document}